# CRAB CAVITIES FOR LINEAR COLLIDERS

G. Burt, P. Ambattu, R. Carter, A. Dexter, I. Tahir, Cockcroft Institute, Lancaster University, Lancaster, UK, LA1 4YR
C. Beard, M. Dykes, P. Goudket, A. Kalinin, L. Ma, P. McIntosh, ASTeC, Daresbury Laboratory, Warrington, Cheshire, WA4 4AD, UK
D. Shulte, CERN, Geneva, Switzerland
R.M. Jones, Cockcroft Institute, Manchester University, Warrington, UK, WA4 4AD
L. Bellantoni, B. Chase, M. Church, T. Khabouline, A. Latina, FNAL, Batavia, Illinois, 60501
C. Adolphsen, Z. Li, A. Seryei, L. Xiao, SLAC, Menlo Park, California

*Abstract*

Crab cavities have been proposed for a wide number of accelerators and interest in crab cavities has recently increased after the successful operation of a pair of crab cavities in KEK-B. In particular crab cavities are required for both the ILC and CLIC linear colliders for bunch alignment. Consideration of bunch structure and size constraints favour a 3.9 GHz superconducting, multi-cell cavity as the solution for ILC, whilst bunch structure and beam-loading considerations suggest an X-band copper travelling wave structure for CLIC. These two cavity solutions are very different in design but share complex design issues. Phase stabilisation, beam loading, wakefields and mode damping are fundamental issues for these crab cavities. Requirements and potential design solutions will be discussed for both colliders.

## INTRODUCTION

Most linear collider concepts envision a crossing angle at the IP to aid the extraction of spent beams. This crossing angle will however reduce the luminosity of the collisions as the beam presents a larger effective transverse size. This loss in luminosity can be recovered by rotating the bunches prior to collision using the time dependant transverse kick of a crab cavity. In particular crab cavities are required for both the CLIC [1] and ILC [2] machines. The proposed solutions for these two colliders are very different and a comparison of the cavities will be the focus of this paper.

A crab cavity is a type of transverse deflecting cavity in which the RF is phased such that the centre of the bunch does not receive a net kick, and the head and tail of the bunch receive equal and opposite kicks [3]. Both travelling and standing wave solutions exist and the cavity can be either normal or superconducting depending on the bunch structure. As the cavity is typically positioned close to the IP before the final doublet their performance can be very sensitive to wakefields. Additionally as the separation between the incoming and extraction beam-lines are very close at this position, the cavities have to be transversely compact.

The voltage, $V_{cav}$, required to cancel the crossing angle of a bunch of energy, $E_0$, is given by equation 1,

$$V_{cav} = \frac{cE_0 \theta_c}{2\omega R_{12}} \quad (1)$$

where $\theta_c$ is the crossing angle, $\omega$ is the cavity frequency and $R_{12}$ is the ratio of the bunch displacement at the IP to the divergence created by the crab cavity. The crab cavity is positioned at a location with a high $R_{12}$ to reduce the required voltage. The ILC has a crossing angle of 14 mrad and an $R_{12}$ of 16.2 m at the crab cavities location. This means a 3.9 GHz system requires a peak deflecting voltage of 2.64 MV at 1 TeV CoM. The CLIC has a crossing angle of 20 mrad and an $R_{12}$ of 25 m; hence a 12 GHz cavity will require a similar voltage of 2.39 MV at 3 TeV CoM.

## PHASE AND AMPLITUDE STABILITY

As both the ILC and CLIC machines have very small transverse bunch sizes at the IP, the phase and amplitude of the crab cavities have to be very stable, as the primary action of a crab cavity is to displace the head and tail of the bunch at the IP. The displacement of a bunch at the IP, $\Delta x$, due to a timing error $\Delta t$ is given by,

$$\Delta x(\Delta t) = R_{12} \frac{V_{cav}}{E_o} \sin(\omega \Delta t) \quad (2)$$

and the luminosity reduction factor, S, is given by

$$S = \exp\left(-\frac{\Delta x^2}{4\sigma_x^2}\right) \quad (3)$$

The horizontal beam size in the ILC is around 500 nm giving a positron-to-electron arm phase tolerance of 80 fs which is around the state of the art level [4]. For the CLIC beam size of 60 nm the timing stability is much smaller at 5 fs which is a major challenge to be overcome and will certainly require all cavities to be driven by a single amplifier.

The amplitude tolerance of a crab cavity is set by the luminosity loss associated with beams colliding with crossing angles. The incorrect amplitude on a crab cavity will cause incorrectly bunch rotation for the crossing angle and the bunches will collide with a small angle between them. The tolerable amplitude stability is given in equation 4

$$\frac{\Delta V}{V_{cav}} = \frac{2}{\theta_c}\frac{\sigma_x}{\sigma_z}\sqrt{\frac{1}{S^2}-1} \qquad (4)$$

This leads to an amplitude tolerance of 4.8 % for the ILC and 2.0 % for the CLIC crab cavities which should not prove difficult to achieve.

## BEAM LOADING

In transverse deflecting cavities the primary action of the RF fields is to kick the bunch transversely. This action has a very small exchange of energy between the electrons and the cavity fields as the electrons gain or loose very little energy. However if the beam traverses the cavity off-axis then the axial electric field component of the dipole fields can accelerate or decelerate the beam in it's direction of motion. This acceleration or deceleration of the bunch in the axial direction causes a large exchange of energy between the bunch and the cavity fields which can alter the amplitude and phase of the cavity fields. As the axial electric field is approximately proportional to the radial offset of the beam, the beam can either give or remove energy from the cavity depending on the exact beam position.

The RF fields induced by the beam have the longitudinal electric field in-phase with the beam, but the transverse voltage is always 90 degrees out of phase with the longitudinal field and hence the beam-loading is out of phase with the peak deflecting field which in turn means that beam-loading fields are in crabbing phase.. This means that crab cavities will have much higher beam loading than deflecting mode cavities which are only loaded by the beams self-field.

## MODAL DISTRIBUTION

The modal pass-band of a dipole cavity is not always sinusoidal due to the coupling between the upper and lower hybrid dipole modal pass-bands [5]. This effect often causes the group velocity to be reduced close to the π mode of the lower (operating) dipole pass-band, depending on the iris radius.

For a standing wave cavity like the design proposed for the ILC, the frequency separation between the π mode and its nearest neighbour is reduced. This can cause interference between these modes, restricting field flatness tuning and LLRF control of the cavity. This limits the number of cells to 9, in order to keep the separation greater than 2 MHz. The ILC cavity modal dispersion diagram is shown in Fig 1.

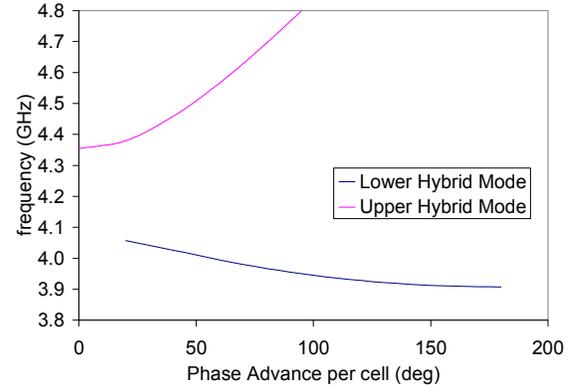

Figure 1: The modal distribution of the first two dipole passbands in the ILC Crab cavity.

For a travelling wave design a low group velocity will increase the effects of beam loading as the energy deposited will take longer to propagate out of the structure. This concern is likely to cause the design to call for the cavity to operate with a phase advance of around π/2 to 2π/3.

## HOM'S, LOM'S AND SOM'S

In any RF cavity there are a number of unwanted modes which may be excited by the beam and must be removed by RF dampers or couplers. In an accelerating cavity where the fundamental mode is the operating mode of the cavity, all the unwanted modes are classed as higher order modes (HOMs), however for a dipole cavity we also have other modes which must be removed. The fundamental mode pass-band of the cavity, which is a lower order mode (LOM), must also be removed to avoid unwanted energy spread. This is not always simple as the LOM is resonant at a lower frequency than the dipole mode and does not penetrate as far down the beampipe as the dipole mode does.

The dipole mode also has two polarisations, a vertical and a horizontal polarisation, which are fixed in place and separated in frequency by using polarising slots, rods or by squashing the cavity. The vertical polarisation of the operating mode, known as the same order mode (SOM) is particularly damaging to the beam due to the small vertical beam sizes and high shunt impedance of this mode (as it will have field shapes close to that of the operating mode).

In the ILC the LOM is damped by the use of a hook type coupler positioned vertically such that it doesn't couple to the operating mode. It is proposed that this coupler could also remove the SOM or a 2nd dedicated co-axial probe could be used. The HOM coupler is a co-axial F-probe type coupler similar to the design used in the ILC main linac [6]. The ILC couplers are shown in Fig 2.

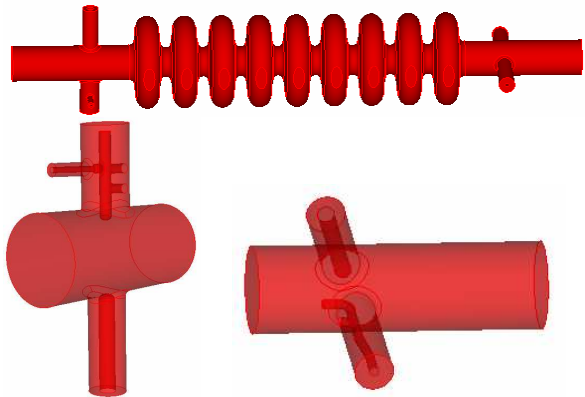

Figure 2: Model of the ILC deflecting mode cavity.

For the CLIC crab cavity a damped-detuned structure is proposed. A number of solutions have been proposed for the damping part; such as manifold damping or choke couplers combined with waveguide couplers at the end of each section. This could be combined with detuning of the SOM to provide very low wakefields for this mode.

## CAVITY FREQUENCY AND TECHNOLOGY CHOICES

For the ILC, a superconducting design was chosen at three times the frequency of the main linac to reduce its size. The size reduction was required to fit a cryostat that did not impinge on the extraction line 21 cm away from the crab cavity central axis. The superconducting design was chosen due to the high duty factor and high gradients required resulted in a very high average power required. The resulting thermal effects in a copper system could possibly cause problems in meeting the phase stability specification [7]. Additionally, the ILC design calls for very lengthy bunch trains and the larger iris of superconducting designs improves the beam-induced wakefield situation.

For the CLIC crab cavity the phase and timing stability requirements are much tighter than for the ILC. As the cavity voltage required decreases with frequency, the phase stability requirement loosens with increasing frequency. In addition it is obvious that with a fixed bunch separation, an increased cavity frequency means more RF periods between bunches. This means that the CLIC cavity should be at as high a frequency as possible. However the crab cavity is placed at the position with the largest beta function in the final focus, which places limits on the aperture size. This lead to the decision to use the main linac frequency of 11.9942 GHz for the crab cavity, which also allows a certain synergy between the two cavity designs [8].

At the frequency and bunch spacing chosen for the CLIC design, a normal conducting cavity is the only viable option.

## CONCLUSION

The crab cavities for the ILC and CLIC colliders are of very different design but they share a number of key similarities that differentiate them from accelerating cavities.

The ILC cavity has completed its design phase and is now moving into a prototyping phase. A single cell Nb prototype and a full 9 cell aluminium prototype including couplers has been fabricated and successfully used to validate the simulations..

The design of a crab cavity suitable for the CLIC collider has commenced and some basic design parameters have been investigated. It is proposed to test a prototype of this cavity at CTF3 in 2012.

## ACKNOWLEDGEMENTS

The authors would like to acknowledge the support and advice of A. Seryei and C. Adolphsen of SLAC. This work was supported by STFC through LC-ABD and the EU through the EUROTeV program.